\documentclass[preprint,showkeys,secnumarabic,amsfonts,showpacs,amsmath,amssymb]{revtex4}
\usepackage[dvips]{color}
\usepackage{array}
\usepackage{rotating}
\usepackage{epsfig}
\usepackage{amsmath}
\usepackage{graphicx}
\begin{document}
\title{ Varying alpha and cosmic acceleration in Brans-Dicke-BSBM theory: stability analysis and observational tests }

\author{H. Farajollahi}
\email{hosseinf@guilan.ac.ir}
\affiliation{$^1$ Department of Physics, University of Guilan, Rasht, Iran}
\affiliation{$^2$ School of Physics, University of New South Wales, Sydney, NSW, 2052, Australia}
\author{A. Salehi}
\email{a.salehi@guilan.ac.ir} \affiliation{Department of Physics,
University of Guilan, Rasht, Iran}

\date{\today}

\begin{abstract}
 \noindent \hspace{0.35cm}

By integration of generalized BSBM and Brans-Dicke cosmological models, in this article, we investigate the theoretical framework of fine structure constant variation and current cosmic acceleration. We first develop a mathematical formalism to analyse the stability of the model. By employing observational data to constrain the model parameters, phase space is performed and the attractor solutions of the model are detected. We then examine the model against observational data such as observational Hubble parameter dataset and quasar absorption spectra. The results confirms current universe acceleration and also predicts fine structure constant variation. Furthermore, extrapolation of the best fitted model in high redshift ($z> 15$) illustrates a significantly larger variation of fine structure constant in earlier epoch of the universe.

\end{abstract}


\maketitle

\section{introduction}

There have been many researches of the cosmological consequences of allowing some of the physical constants of
nature varying in spacetime. One of the first ones is that of varying the Newtonian gravitation
constant, $G$, which leads to the Brans-Dicke (BD) scalar-tensor theory of gravity \cite{brans1}. A second quantity the electron charge, and/or the fine
structure constant, $ \alpha=e^2/\hbar c$, in cosmological time where originally introduced in \cite{Teller}. In general, we have no enough information of why the constants take their current values, or whether they are truly constant or not. Nevertheless, in case of fine structure constant variation, motivated by recent accurate observational evidence from the quasar absorption spectra \cite{Webb}-\cite{Murphyc}, cosmologists seek to find models to justify the variation. Due to the first observational evidence, the fine structure constant might change with time; smaller than its present value by $\frac{\Delta \alpha}{\alpha}\equiv \frac{\alpha -\alpha_0}{\alpha_0}\sim 10^{-5}$ \cite{webb1}-\cite{webb4}. A varying $\alpha$ due to variation of electron charge originally proposed by Bekenstein \cite{beken} and then after the first observational evidence from quasar absorption spectra revived and generalized by Sandvik, Barrow and Magueijo in the so called BSBM proposal \cite{Barrow1}. In BSBM model, an scalar field coupled to the electromagnetic part of the Lagrangian is responsible for the alteration of fine structure constant. Later, a generalized BSBM model has been studied by Barrow in \cite{Barrow2} where the coupling constant $\omega$ in the theory is assumed to be a function of the scalar field in the model.

From a different approach, in \cite{farajollahi2}, we investigated both universe acceleration and fine structure constant variation in BSBM theory in the presence of an exponential self potential for the scalar field in BSBM model. We performed stability analysis and constrained the model with the observational
data. The attractor solutions are detected for the best fitted model parameters. We then checked the model against Hubble parameter dataset and quasar absorption spectra. The numerical computation verified both redshift dependence in the fine structure constant and also current cosmin acceleration.

Barrow et al. \cite{borrow00} also in a different work investigated a cosmological model that incorporates variation of both Newtonian gravitation
constant and fine structure constant. Following \cite{borrow00}, in a new framework in this manuscript, we also integrate both Brans-Dicke (BD) and generalized BSBM theories in which the gravitational constant $G$ is replaced by the BD scalar field $\phi$ and fine structure constant is related to BSBM scalar field $\psi$. In comparison with the work in \cite{borrow00}, here, we assume a generalized coupling constant as a function of scalar field $\omega(\psi)$.  In addition, we constrain the model parameters by using recent observational data of Type Ia Supernovae \cite{Reiss}-\cite{Reiss2} and perform stability and phase space investigation. The stability analysis urges us to find attractor solutions of the dynamical system as asymptotically stable solutions \cite{farajol1}-\cite{farajol6}. We therefore avoid fine tuning issue in which the dynamical trajectories in the phase space lie along a common track despite starting from different initial conditions. We finally check the model against observational data of hubble parameter \cite{Ma} and quasar absorption spectra.

\section{Stability of the model and constraints on the parameters}

We start with the integration of Brans-Dicke and BSBM models. The action describing the dynamics of the system with varying both fine structure constant, $\alpha$ and gravitational constant, $G$, takes the form
\begin{eqnarray}\label{action}
S=\int d^{4}x\sqrt{-g}(\phi R-\omega_{BD}\frac{\phi_{,\mu}\phi^{,\mu}}{\phi}+\frac{16\pi}{c^{4}}{\cal L}),
\end{eqnarray}
where
\begin{eqnarray}\label{L}
{\cal L}=-\frac{\omega(\psi)}{2}\psi_{,\mu}\psi^{,\mu}+{\cal L}_{em} e^{(-2\psi)}+{\cal L}_{m}.
\end{eqnarray}
The fine structure constant, $\alpha$, and gravitational constant, $G$, respectively are given by two different scalar fields;  $\alpha=\alpha_{0}e^{2\psi}$ and $G=1/\phi$. We assume that the parameter $\omega(\psi)=\frac{\hbar c}{l^2}$, as a function of scalar field $\psi$, still determines the strength of coupling between scalar field and photons. The characteristic length scale, $l$, is introduced for dimensional reasons and gives the scale down to which the electric field around a point charge accurately obeys Coulomb force law. From the present experimental constraints, in order to avoid conflict with observation, the corresponding energy scale, $\frac{\hbar c}{l}$, has to lie between a few tens of MeV and the Planck scale $\sim 10^{19}$ GeV. A varying $\omega(\psi)$ is due to variation of length scale, $l$. In the conventional BSBM theory, variation in $\alpha$ is assigned to the electron change $e$. The negative sign of the first term in the Lagrangian density, (\ref{L}), usually represents a phantom field. However, as we see later, our model does not produce a phantom regime in the history of the universe as there also exist other factors contributing to the dynamics of the system. The electromagnetic lagrangian is ${\cal L}_{em}=-\frac{1}{4}f_{\mu\nu}f^{\mu\nu}$ where $f_{\mu\nu}$  is the electromagnetic field tensor. From action (\ref{action}), the field equations in flat FRW cosmologies  are
\begin{eqnarray}
H^{2}&=&\frac{1 }{3\phi}(\rho_{m}(1+|\zeta|e^{-2\psi})+\rho_r e^{-2\psi}+\frac{\omega}{2}\dot{\psi}^{2})-H\frac{\dot{\phi}}{\phi}+\frac{\omega_{BD}}{6}\frac{\dot{\phi^{2}}}{\phi^{2}},\label{fried1}\\
2\dot{H}&+&3H^2=\frac{1}{\phi}(-\frac{1}{3}\rho_r e^{-2\psi}-\frac{\omega}{2}\dot{\psi}^{2})-2H\frac{\dot{\phi}}{\phi}-\frac{\omega_{BD}}{2}\frac{\dot{\phi^{2}}}{\phi^{2}}-\frac{\ddot{\phi}}{\phi},\label{fried2}
\end{eqnarray}
where we put  $8\pi =c=\hbar=1$ and assume a perfect fluid filled the cosmic with $p_{m}=\gamma\rho_{m}$. The ratio $\zeta=\frac{{\cal L}_{em}}{\rho_{m}}$ in equation (\ref{fried1}) describes the fraction of nonrelativistic
matter in the universe that contributes to ${\cal L}_{em}$.
The energy density $\rho_{m}$ stands for the contribution
from cold dark matter (CDM) to the energy density. We also neglect radiation energy density, $\rho_r=0$. In addition, the field equations for the scalar fields $\phi$ and $\psi$ respectively are
\begin{eqnarray}\label{phiequation}
\ddot{\phi}+3H\dot{\phi}=\frac{\rho_{m}(1+|\zeta|e^{-2\psi})}{3+2\omega_{BD}}-\frac{\omega\dot{\psi^{2}}}{6+4\omega_{BD}},
\end{eqnarray}
\begin{eqnarray}\label{phiequation2}
\ddot{\psi}+3H\dot{\psi}+\frac{(d\omega /d\psi)\dot{\psi}^{2}}{2\omega}=2\frac{|\zeta|}{\omega}\rho_{m}e^{-2\psi}.
\end{eqnarray}
To reduce the complexity of the nonlinear field equations (\ref{fried1})-(\ref{phiequation2})and also perform stability analysis and plot phase space trajectories, we introduce the following new variables:
\begin{eqnarray}\label{conserv1}
\chi^{2}=\frac{\rho_{m}}{3\phi H^{2}},\ \ \xi^{2}=\frac{\omega\dot{\psi^{2}}}{\phi H^{2}},\ \ \eta^2=\frac{\dot{\phi}^2}{\phi^2 H^2},\ \ \theta^2=\frac{\phi}{\omega}.
\end{eqnarray}
Employing the Friedmann constraint equation (\ref{fried1}), in terms of new variables, the fine structure constant takes the from:
\begin{eqnarray}\label{alpha1}
\alpha=\chi^2\alpha_{0}|\zeta|(1-\chi^{2}-\frac{\xi^{2}}{6}+\eta-\omega_{BD}\frac{\eta^{2}}{6})^{-1}.
\end{eqnarray}
For further consideration, in term of new variables the $\frac{\dot{H}}{H^2}$ is found to be,
\begin{eqnarray}\label{q}
\frac{\dot{H}}{H^2}=-\frac{3}{2}\chi^{2}(1+\frac{1}{3+2\omega_{BD}}+\frac{|\zeta|\alpha_{0}}{\alpha})+\omega_{BD}\eta^{2}
+\frac{1+\omega_{BD}}{3+2\omega_{BD}}\xi^{2}+2\eta.
\end{eqnarray}
In terms of these variables, the field equations (\ref{fried1})-(\ref{phiequation2}) become,
\begin{eqnarray}
\chi^{'}&=&-\frac{\chi}{2}(\eta+3+2\frac{\dot{H}}{H^2}), \label{conserv2}\\
\xi^{'}&=&-\frac{\xi}{2}(\eta+6+2\frac{\dot{H}}{H^2})-6\frac{|\zeta|\alpha_{0}}{\alpha}\chi^{2}\theta^{2},\label{conserv3}\\
\eta^{'}&=&-\eta(\eta+3+\frac{\dot{H}}{H^2})+\frac{3\chi^{2}(1+\frac{|\zeta|\alpha_{0}}{\alpha})-\xi^{2}}{3+2\omega_{BD}},\label{conserv5}\\
\theta^{'}&=&\theta(\eta-\beta),\label{conserv6}
\end{eqnarray}
where $ ^{'}$ denotes derivative with respect to $ N=\ln (a)$ and $\beta$ is a dimensionless constant in $\omega\propto e^{\beta N}$. Next, we perform stability analysis while constraining the model parameters by using the distance modulus of 557 Type Ia supernovae (SNeIa) from the
Union 2 sample. Note that $\frac{\alpha_{0}}{\alpha}$ and $\frac{\dot{H}}{H^2}$ in the above equations are given by relations (\ref{q}) and (\ref{alpha1}). Therefore, the only parameters in the model are  $\omega_{BD}$, $\beta$, in addition to the initial conditions for the dynamical variables, $\chi$, $\xi$, $\eta$, $\theta$ and Hubble parameter $h_0$. To derive constraints on cosmological parameters we also need the
luminosity distance ($d_L$) given as a function of the above model parameters and initial conditions. From numerical calculation, Table I shows the best-fitted model parameters, initial conditions and $\chi^2$ minimum value.\\

\begin{table}[ht]
\begin{tabular}{|c | c| c| c| c| c| c| c| c|} 
\hline 
parameters  &  $\omega_{BD}$  &  $\beta$ \ & $\chi(0)$\ & $\xi(0)$\ & $\eta(0)$\ & $\theta(0)$ \ & $h_0$ \ & $\chi^2_{min}$\\ [2ex] 
\hline 
&$-1.6$  & $-0.4$ \ & $0.4$\ & $0.2$\ & $-1$\ &  $\pm 2 \times 10^{-6}$ \ & $0.702$ \ & $543.8467$ \\
\hline 
\end{tabular}\\ \ \
Table I: Constraints on model parameters from SNeIa data.
\label{table:1} 
\end{table}\

The system, modeled by the autonomous differential equations (\ref{conserv2})- (\ref{conserv6}), depends on the numerical values of the parameters that appear in the equations. we already constrained these parameters. By performing linear stability analysis, we find seven critical points given in Table II: \\

\begin{table}[ht]
\begin{tabular}{|c | c|} 
\hline 
Critical points  &  $(\chi$, $\xi$, $\eta$, $\theta$) \ \\ [2ex] 
\hline 
$P1$&($0$,  $0$, $0$, $0$)\ \\
\hline 
$P2_{A,B}$&($0$,  $\pm 3 E$, $-6$, $0$)\  \\
\hline 
$P3_{A,B}$&($0$,  $\pm E$,  $-2$,  $0$)\ \\
\hline 
$P4_{A,B}$&($\pm\frac{F}{1+\omega_{BD}}$, $0$, $\frac{1}{1+\omega_{BD}}$, $0$)\ \\
\hline 
\end{tabular}\\
Table II: Characteristics of the critical points\label{table:1} 
\end{table}
where $E=\sqrt{-2(3+2\omega_{BD})}$ and $F=\sqrt{2+\frac{17}{6}\omega_{BD}+3\omega_{BD}^{2})}$. From Tables I and II, all the critical points for the best fitted model parameters are real. The eigenvlaues for the first five critical points are respectively, $(0, -\beta, -3/2, -3/2)$, $(6, -3, 3/2, -\beta-6)$ and $(1, 2, 3/2, -2-\beta)$ where for the best fitted parameter $\beta=-0.4$, these points are unstable. On the other hand, the eigenvalues for critical points $P4_{A,B}$ are $(\frac{-3}{2}, -\frac{3\omega_{BD}+4}{2\omega_{BD}+2}, \frac{1}{2\omega_{BD}+2}, \frac{1-\beta-\omega_{BD}\beta}{\omega_{BD}+1})$. The stability conditions for these two symmetric critical points are $\beta<\frac{1}{1+\omega_{BD}}$ and $\omega_{BD}<-\frac{4}{3}$ where, for the best fitted parameters $\beta$ and $\omega_{BD}$, are stable points. In a multidimensional phase space, a single critical point characterizes the entire system at an instant of time. The set of trajectories in phase space for all possible initial conditions and for a given set of stability parameters forms phase portrait of the system. From stability point of view, for attractors the trajectories rapidly return to the stable points after finite perturbations. The basin of attraction of an attractor is the set of all points for which the trajectories asymptotically approach it. For a physical interpretation, it is more appropriate to analyze the lower dimensional phase space which is formed by the intersection of phase portrait with an hypersurface. In our model, the 4-dim phase space is therefore reduced to a 3-dim one by projecting the phase space $(\chi, \xi, \eta, \theta)$ into the subspace $(\eta, \xi, \chi)$ at $\theta=0$. In Fig.1, qualitatively the behavior of these solutions are exhibited by sketching the phase space portrait for the stable critical point.

\begin{tabular*}{2.5 cm}{cc}
\includegraphics[scale=.4]{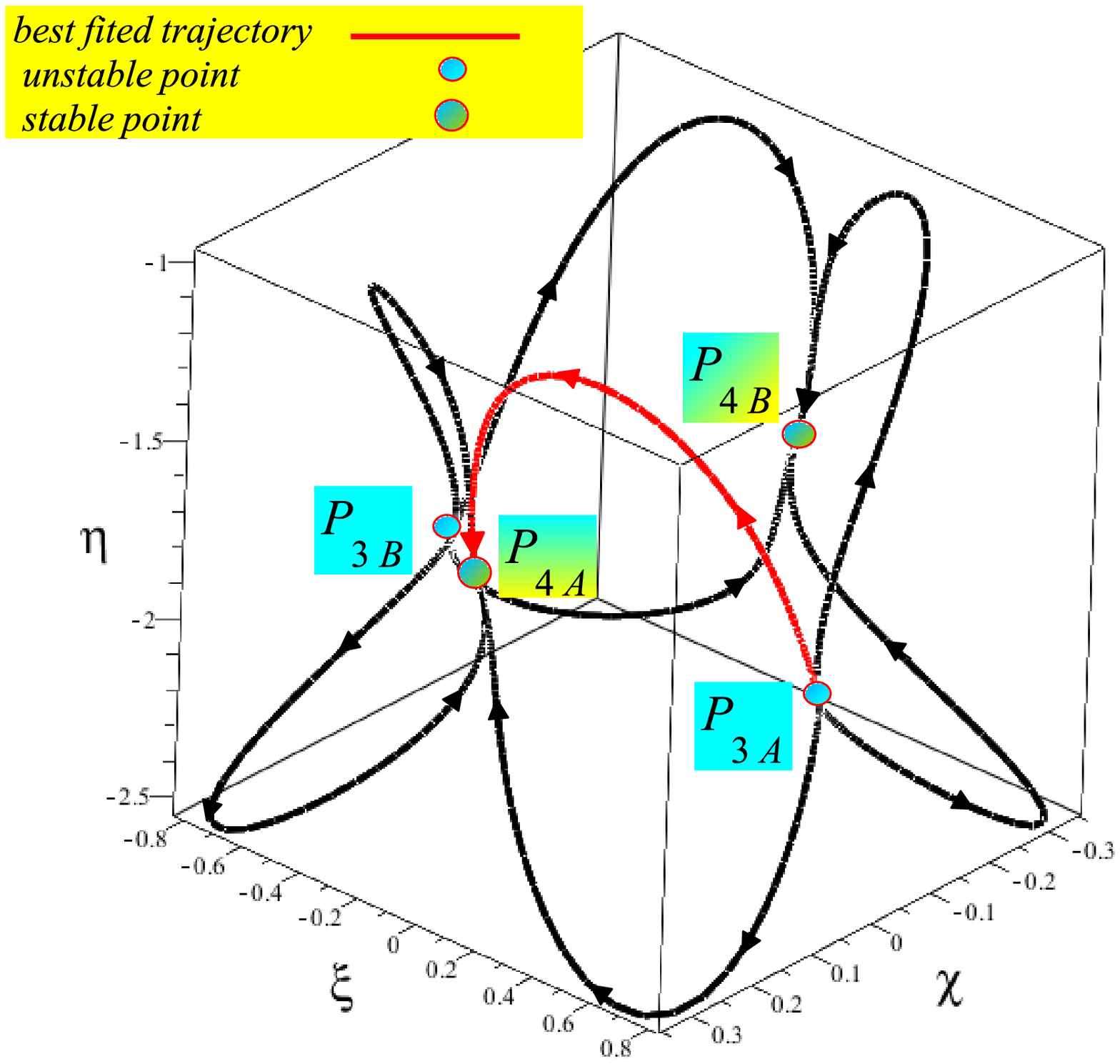}\hspace{0.1 cm}\\
Fig 1:   The projection of 4-dim phase space into 3-dim phase plane for the stable\\ symmetric critical
points $P4_{A,B}$ with the best fitted model parameter. \\
\end{tabular*}\\

The phase space trajectories usually define a set of unstable and stable (attractor) critical points. Following a small perturbation, the trajectories quickly return to the symmetric (with respect to $\chi=0$ line) attractors $P4_{A,B}$. The red trajectory is the best fitted one with the observational data, leaving the unstable critical point $P3_A$ and entering the stable critical point $P4_A$. A more detailed discussion of the physical properties of the system with cosmological application is given in the next section.

\section{Discussion}

In this section, for the model under consideration, we test the dynamical behavior of the cosmological quantities. The reconstructed effective equation of state parameter, $w$ in terms of the phase space dynamical variables given by
\begin{eqnarray}\label{omega}
w&=&(1+\frac{1}{3+2\omega_{BD}}+\frac{|\zeta|\alpha_{0}}{\alpha})\chi^{2}-
\frac{2}{3}\omega_{BD}\eta^{2}+\frac{-2(1+\omega_{BD})}{3(3+2\omega_{BD})}\xi^{2}-\frac{4}{3}\eta-1.
\end{eqnarray}
From numerical calculation, the parameter $w$ is shown in Fig.2. The best fitted trajectory (dashed red line) shows that the universe leaves the unstable state in very high redshift ( radiation dominated era) and enters the current quintessence stable state. The model does not exhibit phantom crossing behavior in the past, but predict that the universe eventually reaches the stable quintessence era at the present era (with $w_0=-0.66$ within the limit of our observations estimation \cite{Rapetti}). The other trajectories correspond to the dynamics of the universe for arbitrary (not fitted) model parameters.\\

\begin{tabular*}{2.5 cm}{cc}
\includegraphics[scale=.4]{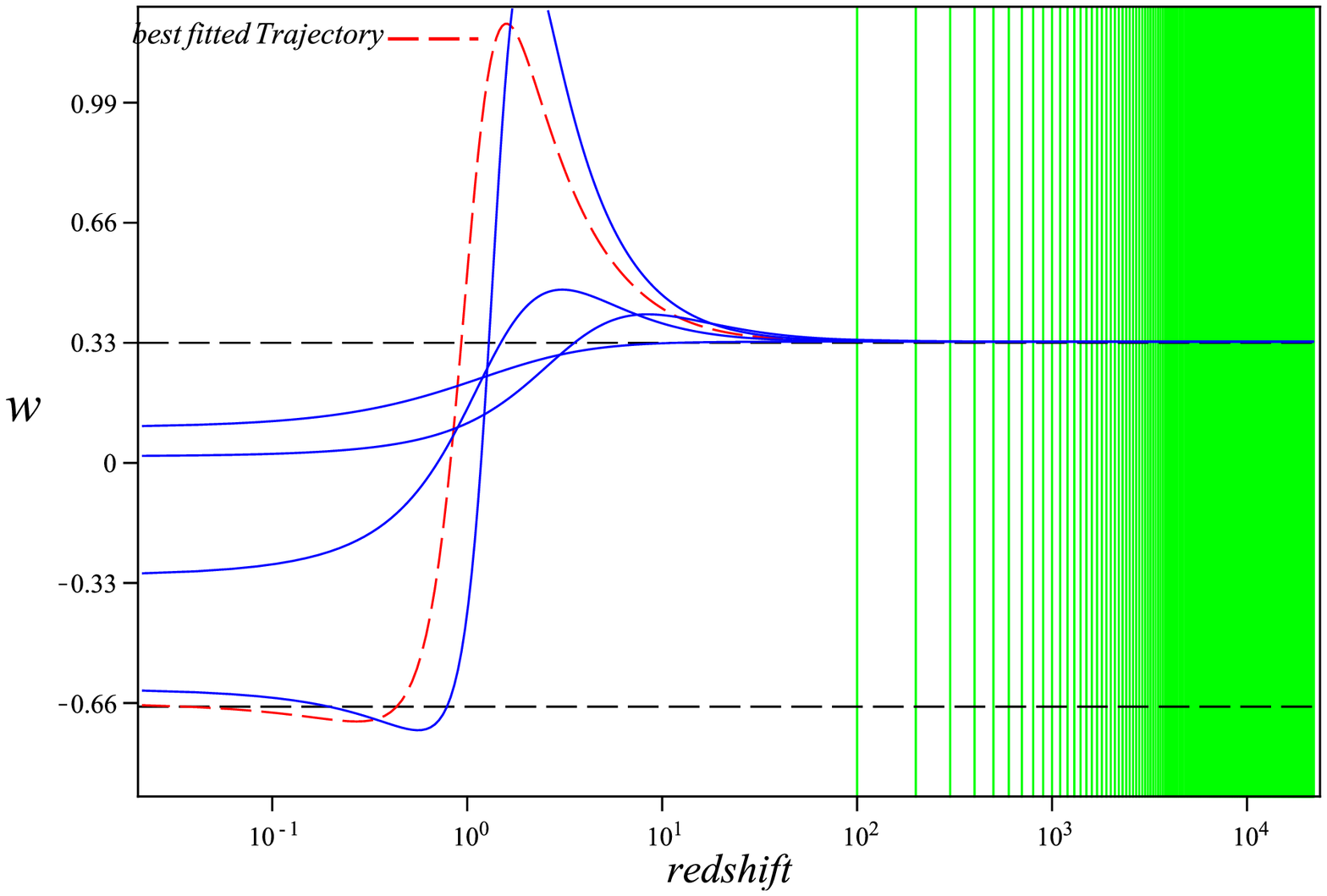}\hspace{0.1 cm}\\
Fig. 2:  The reconstructed equation of state parameter, $w$ , as a function of redshift.\\
The best fitted parameter leaves the unstable state in the past corresponding to \\ the
 critical point $P_{3A}$ and enters the current sable state corresponding to the\\
  stable critical point $P_{4A}$
\end{tabular*}\\

In direct contact with observation the time shift density parameter $\frac{\Delta\alpha}{\alpha}$ is the next quantity to be checked against observation. In terms of new variables, the parameter is given by,
\begin{eqnarray}
\frac{\Delta\alpha}{\alpha}&=&\frac{\frac{|\zeta|\chi^{2}}{2}}{[q+1-\frac{3\chi^{2}+\xi^{2}}{2(3+2\omega_{BD})}
-\frac{3}{2}\chi^{2}+2\eta+\omega_{BD}\eta^{2}]}-1. \label{deltaalpha}
\end{eqnarray}
From numerical calculation, in Fig 3, the parameter is compared with its quasar absorption spectra observational dataset \cite{webb5}. We see that the best fitted model parameter, while passing through the datapoints, is not very sensitive to $\alpha$ variation in $10^{-5}$ scale. The model extrapolation predicts that for redshift $z> 15$ the time shift density parameter begins to vary significantly. Again, for comparison, we also plotted the fine structure constant for arbitrary (not fitted) model parameters.

\begin{tabular*}{2.5 cm}{cc}
\includegraphics[scale=.47]{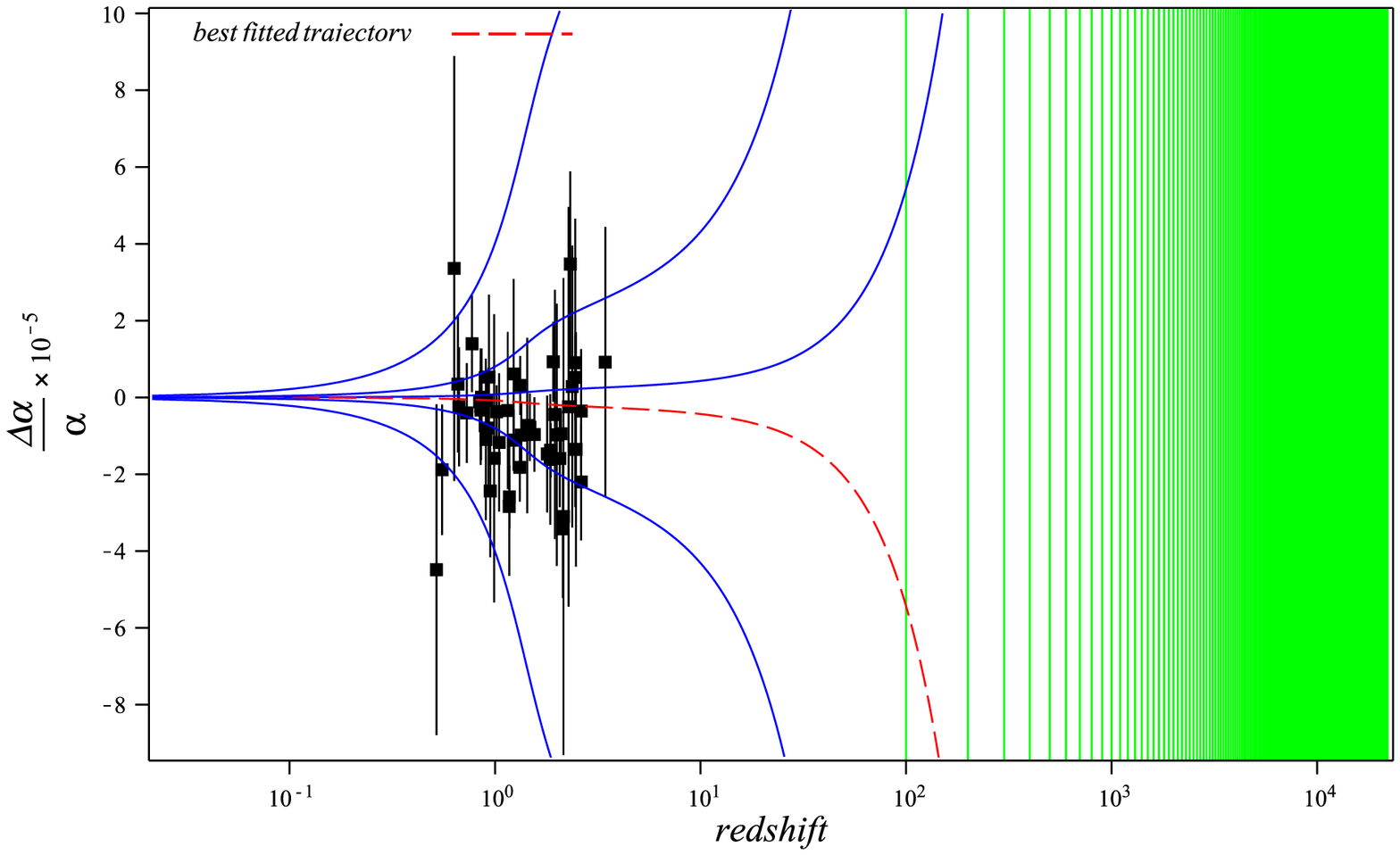}\hspace{0.1 cm}\\
Fig. 3:  The time shift density parameter, $\Delta \alpha/\alpha$, plotted for the model against\\quasar absorption spectra observations. The best fitted trajectory  \\
is shown with red dashed line.
\end{tabular*}\\

Furthermore, we compare the behavior of Hubble parameter derived from numerical calculation in the model with the observational data \cite{hubbledata}. From Fig. 4 the best fitted Hubble parameter in the model agrees with the data.

\begin{tabular*}{2.5 cm}{cc}
\includegraphics[scale=.4]{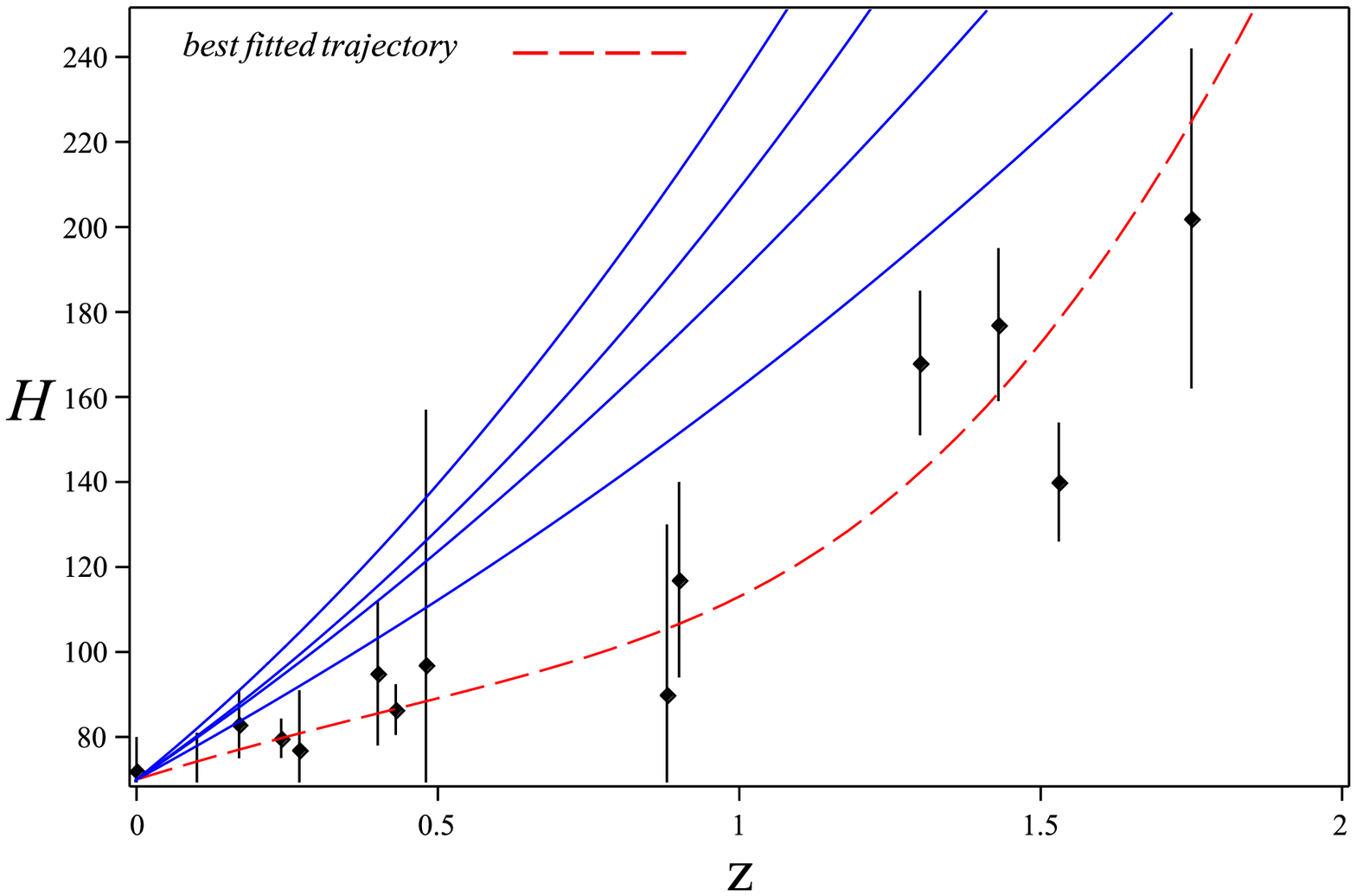}\hspace{0.1 cm}\\
Fig. 4:  The graph of derived Hubble parameter $H$ as a function of redshift \\in comparison with the data. The best fitted parameters is shown with a red \\dashed curve. \\
\end{tabular*}\\

Finally, the reconstruct scalar field responsible for both $\alpha$ variation and universe acceleration using the best fitted model parameters is shown in Fig. 5.

\begin{tabular*}{2.5 cm}{cc}
\includegraphics[scale=.4]{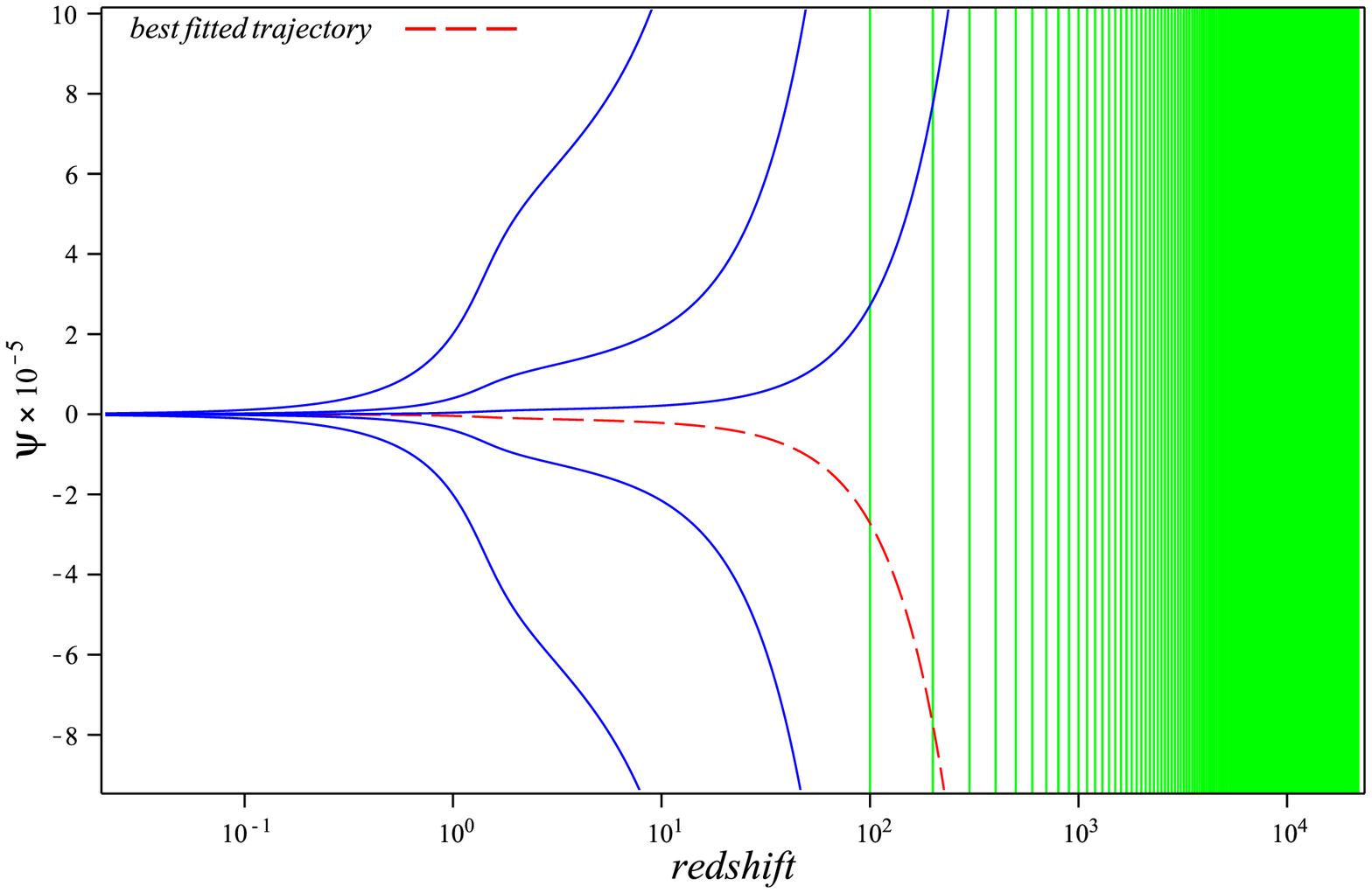}\hspace{0.1 cm}\\
Fig. 5:  The graph of the scalar field $\psi$ as a function of redshift. The best fitted scalar field\\ with the model parameters is shown with a red dashed curve. \\
\end{tabular*}\\

\section{Summary and Remarks}

In this paper we have introduced a criterion that can be used to probe the cosmological
viability of gravitational and fine structure constants ( $G$, and $\alpha$) variation theories. Specifically, we investigated the integration of BD and BSBM theories where the two BD and BSBM scalar fields are responsible both for the universe acceleration and $\alpha$ variation. In practice, the model has simultaneously satisfied both observational evidence from SNeIa and quasar absorption spectra. We first best fitted the model with the observational data from SNeIa for distance modulus and found constraints on the model parameters. We then performed two quantitative and qualitative analyses. The validity of model with the constraints on its parameters are verified by these analyses. Qualitatively, we performed stability analysis and found the attractor solutions in the phase space. These attractor solutions start with a small perturbation of the model parameters in the radiation dominated epoch and exhibit the final resting state of the universe in an accelerated expansion phase. The matter dominated phase of the universe is just a transient state in this model. Two quantitative tests are also performed to directly compare the model with the observational data; the observational Hubble parameter and $\alpha$ variation tests. The best fitted model with the attractor property are verified by these two tests. The model also predicts that variation of fine structure constant become greater in higher redshift. From equation of state parameter the universe is in quintessence regime at the present time. This is in consistent with the attractor property of the equilibrium solution in a scalar field dominated cosmology. This is an advantage over those cosmological models that predict universe acceleration in phantom era where the null energy is violated. In addition, by extrapolating backward in time the model predicts a significantly larger variation of fine structure constant in earlier epoch of the universe.

\end{document}